\documentclass[rmp]{revtex4}

\usepackage{graphicx}
\usepackage{amsmath}
\usepackage{natbib}

\newcommand{\EQ}[1]{Eq.~(\ref{eq:#1})}

\newcommand{\beast}[1]{\emph{#1}}
\newcommand{\xs}{\chi}
\newcommand{\ys}{\eta}
\newcommand{\ws}{\tilde{w}}
\newcommand{\rs}{\tilde{r}}
\newcommand{\dss}{\tilde{s}}
\newcommand{\X}{X}
\newcommand{\Y}{Y}
\newcommand{\x}{x}
\newcommand{\y}{y}
\newcommand{\w}{w} 
\newcommand{\rec}{r}
\newcommand{\ds}{s}
\newcommand{\mut}{U_b}

\newcommand{\pfix}{P_{e}}
\newcommand{\psfix}{p_{e}}

\newcommand{\fitdis}{P}
\newcommand{\be}{\begin{equation}}
\newcommand{\ee}{\end{equation}}
\newcommand{\bea}{\begin{eqnarray}}
\newcommand{\eea}{\end{eqnarray}}
\newcommand{\ba}[1]{\begin{array}{*{#1}{c}}}
\newcommand{\ea}{\end{array}}

\newcommand{\B}{\Theta}
\newcommand{\lpvar}{\rho}

\newcommand{\OM}{\Omega}
\newcommand{\stp}{\sqrt{2\pi}}
\newcommand{\stps}{\sqrt{2\pi\sigma^2}}

\begin{document}

\title{Rate of Adaptation in Large Sexual Populations}
\author{R.~A.~Neher${}^{*}$}
\author{B.~I.~Shraiman${}^{*\ddagger}$}
\author{D.~S.~Fisher${}^{\dagger}$}
\affiliation{${}^{*}$Kavli Institute for Theoretical Physics}
\affiliation{${}^{\ddagger}$Department of Physics, University of
California, Santa Barbara, CA 91306}
\affiliation{${}^{\dagger}$Department of Applied Physics, Stanford University,
Stanford, CA 94305}
\date{\today}

\begin{abstract}
Adaptation often involves the acquisition of a large number of genomic changes
which arise as mutations in single individuals. In asexual populations,
combinations of mutations can fix only when they arise in the same lineage, but
for populations in which  genetic information is exchanged, beneficial mutations
can arise in different individuals and  be combined later. In large populations,
when  the product of the population size $N$ and the total beneficial mutation
rate $\mut$  is large, many new beneficial alleles can be segregating in the
population simultaneously.  We calculate the rate of adaptation, $v$, in several
models of such sexual populations and show that $v$ is linear in $N\mut$ only in
sufficiently small populations. In large
populations, $v$ increases much more slowly as $\log N\mut$. The prefactor of
this logarithm, however, increases as the square of the recombination rate.  
This acceleration of adaptation by recombination implies a strong evolutionary
advantage  of sex.
\end{abstract}
\maketitle

In asexual populations, beneficial mutations arising on different genotypes
compete against each other and in large populations most of the beneficial
mutations are lost because they arise on mediocre genetic backgrounds, or acquire
further beneficial mutations less rapidly than their peers --- the combined
effects of clonal interference and multiple
mutations \cite{Gerrish_Genetica_1998,Desai_Genetics_2007}. Exchange of genetic
material between individuals allows the combination of beneficial variants which
arose in different lineages,  and can thereby speed up the process of adaptation
\cite{Fisher_1930,Muller_AmericanNaturalist_1932}. Indeed, most life forms engage
in some form of recombination, e.g. lateral gene transfer or competence for
picking up DNA in bacteria, facultative sexual reproduction in yeast and plants,
or obligate sexual reproduction in most animals. Some benefits of recombination
for the rate of  adaptation have recently been demonstrated experimentally in
\beast{C.reinhardtii} \cite{Colegrave:2002p6888},
\beast{E.coli} \cite{Cooper_PlosBiology_2007}, and \beast{S.cerevisiae}
\cite{Goddard_Nature_2005}, for a review of older experiments see \cite{Rice:2002p28396}.

Yet the benefits of sex become less obvious when one considers its
disadvantageous effects: recombination can separate well adapted combinations of
alleles and  sexual reproduction is more costly than asexual reproduction due to
resources spent for mating and, in some cases, the necessity of males. The latter
--- in animals often termed the two-fold cost of sex ---  implies that sexual
populations can be unstable to the invasion of asexual variants. As a result, the
pros and cons of sex have been the subject of many decades of debate in the
theoretical literature
\cite{Crow_AmericanNaturalist_1965,Smith_AmericanNaturalist_1968,Felsenstein:1974p23937,Barton_GenetRes_1995,Barton_Science_1998},
and several different potentially beneficial aspects of sex have been identified
including the pruning of detrimental mutations
\cite{Peck:1994p29075,Rice:1998p28409} and host-parasite coevolution or otherwise
changing environments
\cite{Ladle:1993p24054,Burger:1999p21459,Waxman:1999p24933,Charlesworth:1993p25080,Gandon:2007p2307,Callahan_2009}.
In the opposite situation of relatively static populations, it has been proposed
that recombination is favored in the presence of negative epistasis
\cite{Kondrashov:1984p28365,Kondrashov:1988p18001,Feldman:1980p28377} - a
situation when the combined detrimental effect of two unfavorable alleles is
greater than the sum of the individual effects. While this may sometimes be  a
significant effect, most populations, especially microbes, are likely to be under
continuing selection and the benefits of sex for speeding up adaptation are
likely to dominate.

The Fisher-Muller hypothesis is that sex speeds up adaptation by combining
beneficial variants.  Moreover, it has been demonstrated by
\citet{Hill_GenetRes_1966} that linkage decreases the efficacy of
selection. This detrimental effect of linkage, known as the ``Hill-Robertson
effect'', causes selection for higher recombination rates, which has been shown 
by analyzing recombination modifier alleles at a locus linked to two competing
segregating loci \cite{Barton:2005p982,Iles:2003p28305,Martin:2006p2722,Otto_Genetics_1997,Roze_Genetics_2006}.
Hitchhiking of the allele that increases the recombination rates with the
sweeping linked loci results in effective selection for increased recombination.

Experiments and simulation studies suggest that the Hill-Roberston effect is more
pronounced and selection for recombination modifiers is stronger in large
populations with many sweeping loci
\cite{Colegrave:2002p6888,Iles:2003p28305,Felsenstein:1974p23937}. However, the
quantitative understanding of the effect of recombination in large populaltions
is limited. \citeauthor{Rouzine_Genetics_2005} have studied the role of
recombination in the context of evolution of drug resistance in HIV finding that
recombination of standing variation speeds up adaptation by producing anomalously
fit  individuals at the high fitness edge of the distribution
\cite{Rouzine_Genetics_2005,GheorghiuSvirschevski:2007p17402}. The effects of
epistatic interactions between polymorphisms and recombination on the
dynamics of selection have recently been analyzed by \citet{Neher:2009p22302}.
Yet none of these works consider the effects of new beneficial mutations. 
In the absence of new mutations (and in the absence of heterozygous
advantage which can maintain polymorphisms)  the fitness soon saturates as most
alleles become extinct and standing variation disappears. Thus the crucial
point which must be addressed is {\it the balance between selection and
recombination of existing variation and the injection of additional variation by new
mutations.}

Here, we study the dynamics of continual evolution via new mutations, selection,
and recombination using several models of recombination. Our primary models most
naturally apply when periods of asexual reproduction occur between matings, so
that they approximate the life style of facultatively outcrossing species such as
\beast{S.~cerevisiae}, some plants,  and \beast{C.~elegans}, which reproduce
asexually most of the time but undergo extensive recombination when outcrossing.
The models enable us to study analytically the explicit dependence of the rate of
adaptation and of the dynamics of the beneficial alleles on the important
parameters such as the outcrossing rate and population size. In an independent
study \citeauthor{Barton_2009} (personal communication) calculate the rate of
adaptation for \emph{obligate} sexual organisms using several different
multilocus models of recombination, including the free recombination model
studied here. The relation of our  work to theirs, and well as to that of
\citeauthor{Cohen_PhysRevLett_2005} \cite{Cohen_PhysRevLett_2005,Cohen:2006p5005} 
who have also studied the effects of recombination with multiple new mutations,
is commented on in the Discussion section.

When deleterious mutations can be neglected, the rate of adaptation is the
product of the rate of production of favorable mutations $N\mut$ ($N$ being the
population size and $\mut$ the genome wide beneficial mutation rate), the
magnitude of their effect, and their fixation probability. The fixation
probability is dominated by  the probability that the allele becomes {\it
established}:  i.e. that it rises to high enough numbers in the population that
it is very unlikely to die out by further stochastic fluctuations. In a
homogeneous population a single beneficial mutation with selective advantage
$\ds$ has a probability of establishment and eventual fixation of
$\pfix\approx\frac{\ds}{1+\ds}\approx \ds$ \footnote{In discrete generation
models, $\pfix\approx2\ds$}\cite{Moran:1959p28969}.
In a heterogeneous population, however, a novel beneficial mutation can arise on
different genetic backgrounds and its establishment probability will thus vary,
being greater if it arises in a well adapted individual. But even well adapted
genotypes soon fall behind due to sweeps of other beneficial mutations and
combinations. In order to avoid extinction, descendants of the novel mutation
thus have move to fitter genetic backgrounds via recombination in outcrossing
events \citep{Rice:2002p28396}. As a result the establishment probability
decreases as the rate of average fitness gain, $v$, in the population increases.
But the rate of average fitness gain, or equivalently, the rate of adaptation
itself depends on the establishment probability. These two quantities therefore
have to be determined self-consistently.

In this paper we analyze several models via self-consistent calculations of the fixation probability of new mutations.
For a given production rate of beneficial mutations $N\mut$, we find that
interference between mutations is of minor importance if the recombination rate
$\rec$ exceeds $\sqrt{4\ds^2N\mut}$. In this regimes, the rate of adaption is $v\approx
N\mut \ds^2$ as found for sequential mutations or in the absence of linkage.
At recombination rates below $\sqrt{\ds^2N\mut/\log N\mut}$, however, $v$ grows only
logarithmically with $\log N\mut$. We find this behavior in all our
models and argue that it obtains more generally. The prefactor of the $\log N\mut$ 
increases with the square of the recombination rate, implying a strong benefit
of recombination in large populations.

\section{Models}
We consider a population of haploid individuals with  fitness (growth rate),
$\X$, determined by the additive effects of a large number of loci each of which
makes small contributions to the fitness. We assume selection is weak enough for the
population dynamics to be described by a continuous time approximation, that 
the population size, $N$, is large enough that $N\ds\gg 1$, and that a wide spectrum of
fitnesses is present, characterized by the fitness variance, $\sigma^2$, of 
the population.
Individuals divide stochastically with a Poisson rate $1+\X-\bar{\X}(t)$, where
$\bar{\X}(t)$ is the mean fitness in the population, and they die, also
stochastically, with rate $1$ (that is, we  use the death rate to set the unit
of time and assume for convenience that $\X-\bar{\X}(t) \ll1$). In addition to 
this asexual growth, individuals outcross with rate $\rec$. Within our models,
outcrossing is an independent process decoupled from division 
(but this does not substantively affect our results). 

The primary model of mating that we study is {\it free recombination}. In an
outcrossing event two randomly chosen parents are replaced by two offspring and
each parental allele is assigned at random to one or the other of the two
offspring. This would be exactly correct if all loci were on different
chromosomes, and can be a reasonable approximation when the number of crossover
sites is large so pairs of substantially polymorphic loci are likely to be
unlinked at each mating.  At the end, we discuss briefly what happens when this
approximation breaks down. When the number of polymorphic loci is large and their
contributions to $\X$ are of comparable magnitude, the distribution of offspring
fitness is well described by a Gaussian distributed around the value midway
between the fitnesses of the two parents, and with  variance $\sigma^2/2$ if loci
are uncorrelated \citep{Bulmer_1980}: this is less than the $\sigma^2$ variance
of the parental population. Note that $\sigma^2$ is proportional to the number of
segregating alleles and represents the extent of genetic variation in the
adapting population. It is not a fixed parameter of the model, but is to be
calculated self-consistently as a function of the population size and  the
mutation and out-crossing rates.
 
In addition to the free recombination model described above, we  study two other
models.  The first is a grossly simplified model of recombination in which a
randomly chosen individual is replaced by an individual whose genome is assembled
by choosing the alleles at each locus according to the allele frequencies in the
{\it entire} population, independent of the ``parents"   (see also
\citep{Barton_2009}). In this case recombinant offspring have fitness
distribution identical to the population distribution. It turns out that this
\emph{communal recombination model}, even if unrealistic, behaves similarly to
the free recombination model  while being much easier to analyze mathematically:
this makes it a good source of insight as well as supporting the contention that
the form of our results is more general than the particular models.

The free recombination model, and even more so the communal recombination model,
overestimate the amount of gene reassortment during outcrossing events by
assuming that all loci are simultaneously unlinked by recombination to the same
extent, independent of their locations on the chromosomes.  To study the
effects of more persistent genetic linkage, we also study a third model in which
only a single locus is exchanged with a mating partner in an outcrossing event,
or --- equivalently --- is picked up from DNA in the environment and randomly
replaces the initial allele at the same locus. This model is reminiscent of
lateral gene transfer among bacteria and related to, but not the same as, the
model studied by
\citet{Cohen_PhysRevLett_2005}.
While this \emph{minimal recombination model}  preserves the linkage of all but one
locus at a time, each locus is equally strongly linked to all other loci. Thus
this model does not approximate the position-dependent crossing-over of chromosomes.

The recombination processes in each of these models are characterized by a rate,
$\rec$, and a function, $K(\X,\Y,t)$, which is the distribution of offspring
fitness $\Y$, given a parent with fitness $\X$ mated with a random member of 
the population. Being the distribution of offspring fitness, the recombination
`kernel' is normalized $\int d\Y K(\X,\Y,t)=1$. Furthermore, since we ignore
epistasis and assume that loci at imtermediate frequencies are in linkage
equilibrium, recombination leaves the fitness distribution $\fitdis(\X,t)\
d\X$ of the population invariant $\int d\X K(\X,\Y,t)\fitdis(\X,t)=\fitdis(\Y,t)$. 
Within the free recombination model, each outcrossing event replaces two parents with two offspring.
However, when following a rare allele, we can focus on the lineage
containing this allele and ignore the fate of the other offspring. Matings
between two individuals with the same rare allele are very infrequent and 
can be neglected. Since we are interested in the effects of recombination, 
we will primarily focus on the limit $\rec \gg \ds$.

\subsection{Branching process and establishment probability}
\begin{figure}[htp]
\begin{center}
	\includegraphics[width=0.7\columnwidth]{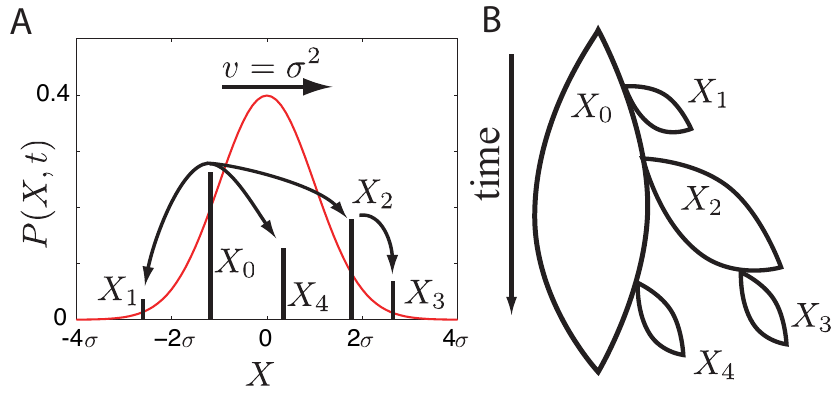}	
  \caption{A novel mutation needs to recombine onto fitter genetic backgrounds
  to become established and eventually fix. Panel A: The distribution in fitness
  of the population moves towards higher fitness with velocity $v=\sigma^2$. The new
  mutation, illustrated by the black bars, has to switch backgrounds by
  recombination to keep up with the moving wave of the population fitness distribution. 
  Panel B: Initially, the novel mutation is present on a single genetic
  background with fitness $\X_0$, struggling not to go extinct. Recombination
  can transfer the mutated allele onto a new background, e.g.~from $\X_0$ to
  $\X_1$, and spawn a daughter clone which starts an independent struggle
  against extinction. The mutation establishes if at least one branch survives indefinitely. The figure shows the
  complementary case of an unsuccessful mutation: all branches die out. The
  probability of establishment, $\w(\X, t)$, depends on the fitness $\X$ of the
  genome in which the mutation arose and is a solution to Eq.~(\ref{eq:non_extinct}).}
  \label{fig:surf_illustration}
\end{center}
\end{figure}

The key element determining the rate of adaptation is the probability that a new
beneficial mutation avoids extinction  and establishes in the population. The establishment probability
is the probability that the allele  survives random drift and rises to a sufficiently large number so that its frequency in the population  grows deterministically (and eventually
fixates). This establishment occurs --- if it does at all --- when the population of the
allele is large but its frequency in the population is still small. The
fate of a new allele during the stochastic phase, when it exists only in a small
fraction of individuals, can be described well by a branching process which
accounts for stochastic birth, death, and, crucially, for recombination events
that move some of  its descendants from one genetic background to another.  The branching
process takes place in a population whose mean fitness is steadily increasing
due to beneficial mutations sweeping and fixing at other loci and in other lineages.
Ignoring the short term effect of mutations, the mean fitness, $\bar{\X} (t)$,  
increases with rate  $v\equiv \frac{d\bar{\X}(t)}{dt} = \sigma^2$, where $\sigma^2$ is the (additive) variance
of the fitness. The dynamics of a novel beneficial mutation linked to a spectrum of
genomic backgrounds in an population adapting with rate $v$ is illustrated in
figure \ref{fig:surf_illustration}. To establish, its descendents have to
switch repeatedly to fitter genomic backgrounds. This general idea (see 
\cite{Rice:2002p28396}  for review) applies to the accumulation of beneficial as
well as deleterious mutations.

The establishment probability at a time $t-dt$ of  descendants of a genome of
fitness $X$, defined as $\w(\X,t-dt)$, is simply related to that at time
$t$ \cite{Barton_Genetics_1995}:
\begin{equation}
\begin{split}
\label{eq:fixation_probability}
\w(\X,t-dt)= &w(\X,t) -dt[D+B(X,t)+\rec]\w(\X,t) +
dt B(\X,t)(2\w(\X,t)-\w(\X,t)^2)\\ &+ dt\ \rec\int d\Y K(\X,\Y,t)\w(\Y,t) 
\end{split}
\end{equation}
where $D=1$ is the death rate and $B(\X)=1+\X-\bar{\X}(t)$ the birth rate. 
After a division, either of the two offspring has a probability $1-\w$ of extinction:
hence $2\w-\w^2$ of at least one of these offspring fixing. For a low-frequency allele conferring additional
fitness $\ds$ on a genomic background with fitness $\X$, we have
$B=1+\X-\bar{\X}(t)+\ds$.

In a sufficiently large population the adaptation process will proceed in a
steady manner leading to a fitness distribution of constant width translating
towards higher fitness as a ``traveling wave'' \citep{Tsimring:1996p19688} 
with the velocity set
by the rate of increase of the mean fitness $v=\frac{d}{dt} {\bar \X}(t) $. 
We make the {\it Ansatz}  that the distribution of fitnesses of  the population
around its mean $\bar{\X}(t)$ does not fluctuate substantially and  
that {\it the distribution is  close to gaussian}. These are analogous to 
``mean-field" approximations which must be justified {\it a posteriori}. 
We expect that such approximations will become valid for sufficiently 
large populations, but how this occurs and how large the population must 
be, is not clear {\it a priori}: we discuss this below.

In the traveling wave population, the establishment probability depends on time
only via ${\bar \X}(t)$. Hence we measure fitness relative
 to ${\bar \X}(t)=vt$, defining $x\equiv \X-{\bar \X}(t)$,
and seek an otherwise time-independent solution of the form
$\w(\x)=\w(\X-vt)=\w(x,t)$. (The properties of $\w(\X,t)$ and $K(\X,\Y,t)$
do not change by this shift of variables other than becoming time independent 
relative to a moving
reference $\bar{\X}(t)$. We therefore use the same symbols for $\w(\x)$ and
$K(\x,\y)$ in the moving frame.) Using $\partial_t \w(X-vt)=-v\partial_x
\w(\x)$, the establishment probability, $\w(\x)$, then obeys
\begin{equation}
\label{eq:non_extinct}
\begin{split}
v\partial_\x \w(\x)
=\rec\int dy K(\x,\y)\w(\y)+(\x+\ds-\rec)\w(\x)-(1+\x+\ds)\w(\x)^2\ .
\end{split}
\end{equation}
In many cases of interest, selection is only important on timescales much
longer than the generation time. In that case $\x+\ds$ in the prefactor 
of the quadratic term is negligible compared to the inverse generation time,
which is $1$ in our units. Eq.~(\ref{eq:non_extinct}) then simplifies to
\begin{equation}
\label{eq:fixprob_rescaled}
\begin{split}
(v\partial_\x - \x+\rec ) \w(\x) - \rec\int d\y K(\x,\y)
\w(\y) \ = \ \ds \w(\x)-\w(\x)^2,
\end{split}
\end{equation}
We have written this in a suggestive form. The left hand side of
Eq.~(\ref{eq:fixprob_rescaled}) defines the linear operator ${\cal J}$ acting on
$\w(\y)$. At very high recombination rates, we will obtain that $\w(\x) \sim
(1+2\x/\rec)$ which is almost independent of $\x$ for $\x\ll\rec$. In this limit, the
${\cal J}$ acting on $\w(y)$ vanishes and the population
average establishment probability is just the solution to the right-hand side, 
giving simply $\w(\x) \approx \ds$. This is the conventional result 
(obtained by the simple branching process) in the absence of
linkage to the rest of the genome. More generally, the fixation probability of a
new mutation which can arise in any individual is  the population average of the
$\x$-dependent establishment probability over the approximately gaussian
distribution of the fitness, $\x$:
\begin{equation}
\label{eq:fixp}
\pfix \approx \int \frac{d\x}{\sqrt{ 2 \pi \sigma^2}}
e^{-\frac{\x^2}{2\sigma^2}} \w(\x)
\end{equation}
Equation~(\ref{eq:fixprob_rescaled}) has an important property. Its left hand
side is {\it zero upon averaging} with respect to the population distribution
$\fitdis(\x)=\frac{1}{\stps}e^{-\x^2/2\sigma^2}$ (as is readily confirmed by
direct integration using $v=\sigma^2$ and $\int d\x
K(\x,\y)\fitdis(\x)=\fitdis(\y)$, see above). This property originates from the fact
that in the deterministic limit (without the additional mutation, $\ds$), 
the population dynamics has $\fitdis(\X,t) = \fitdis(\X-vt) = \fitdis(\x)$ as a
traveling wave solution \citep{Rouzine_Genetics_2005} --- the initial rationale for 
assuming a gaussian form. As a consequence, averaging
Eq.~(\ref{eq:fixprob_rescaled}) yields a ``solvability condition"
\begin{equation}
\label{eq:solv_cond}
\int \frac{d\x}{\sqrt{ 2 \pi \sigma^2}}
e^{-\frac{\x^2}{2\sigma^2}}\left(\ds\w(\x)-\w(\x)^2\right)=0
\end{equation}
which, when combined with Eq.~(\ref{eq:fixp}), provides another expression for the 
establishment probability:  
\be
 \ds \pfix = \int \frac{d\x}{\sqrt{ 2 \pi \sigma^2}}
e^{- \x^2 /2\sigma^2 } \w(\x)^2\ .
\ee

This equation together with Eq.~(\ref{eq:fixprob_rescaled})
describes the ``surfing''  of a beneficial allele (and far more often its
drowning!) --- the processes illustrated by figure \ref{fig:surf_illustration} 
--- under the assumption that the distribution of fitness in the
population is sufficiently close to gaussian. The latter holds when the large number of alleles
at different loci  are only weakly correlated: we justify this {\it Ansatz}
below.

\subsection{Models of recombination}
The recombination kernel $K(\x,\y)$ depends on the recombination model.  For 
the {\it free recombination model,} the fitness of the offspring resulting
from a mating of two parents with fitness $\x$ and $z$ is again Gaussian
distributed with mean $(\x+z)/2$ and variance $\sigma^2/2$. Averaging over 
the fitness $z$ of the mate, which is Gaussian distributed with variance
$\sigma^2$, results in the recombination kernel
\begin{equation}
\label{eq:reckernel}
K(\x,\y)=\sqrt{{2 \over 3\pi\sigma^2}} e^{- \ {2(y- {x \over 2} )^2 \over
3\sigma^2}}\ .
\end{equation}

In the {\it communal recombination model}, the fitness of the recombinant  is a
random sample from the population (assuming gaussianity and linkage equilibrium).
In that case, we have 
\begin{equation}
K(\x,\y)=\frac{1}{\sqrt{2\pi\sigma^2}}e^{-\frac{\y^2}{2\sigma^2}},
\end{equation}
i.e. the recombination kernel becomes independent of $x$ and equation
\EQ{fixprob_rescaled} becomes mathematically much simpler.

Within the {\it minimal recombination model}, the probability per unit time of
any particular locus being transferred is $\rec$ and the sections are assumed
small enough that they contain at most one segregating locus. From the point of view of
a single mutant, there are two processes: either it can be transfered to another
genome, which is effectively like the recombination process in the communal
recombination model, or other sections can be transfered into its genome
gradually changing its fitness. With small sections transfered the fitness of the
genome undergoes a random walk with bias towards the average fitness. The
corresponding recombination operator is then 
\be
\label{eq:minimalreckernel}
\rec\int d\y K(\x,\y) \w(\y)= \rec\int \frac{
d\y}{\sqrt{2\pi\sigma^2}}e^{-\frac{\y^2}{2\sigma^2}}\w(\y) + \rec[\sigma^2
\frac{d^2\w}{d\x^2}-\x \frac{d\w}{d\x}] \ .
\ee
This form of the recombination operator is derived in the Appendix \ref{sec:AppMinimalRec}.
Note that for the minimal recombination model the recombination operator
acting on $\fitdis(\x)$ is different from the adjoint operator acting on
$\w(\y)$.

\section{Results}
\subsection{Fixation probability and rate of adaption}
To calculate the rate of adaptation, we solved Eq.~(\ref{eq:fixprob_rescaled})
and obtained expressions for the average fixation probability $\pfix$ of a
beneficial mutation, which is of the form $\pfix=\sigma \psfix(\rs,\dss)$, where
$\dss=\ds/\sigma$ and $\rs=\rec/\sigma$ are the selective advantage of the
beneficial mutation and the outcrossing rate rescaled by the to-be-determined
width of the fitness distribution $\sigma$. The expression for $\pfix$ is used
later to calculate $\sigma^2$ in a self-consistent manner. The derivation of the
expressions for $\pfix$ in the different models are given in the following
section. In the limit $\ds\ll\rec$, our primary focus, we find for the free
recombination model
\begin{equation}
\label{eq:establishmentprob}
\psfix(\rec/\sigma, \ds/\sigma) = 
\begin{cases}
\vspace{0.2cm}
\frac{\sigma^2\log(c\rec/\ds)}{\ds\rec\stp}e^{-\frac{\sigma^2}{2\rec^2}\log^2(c\rec/\ds)}
& \ds\ll\rec \ll \sigma \\ 
\frac{\ds}{\sigma}\left(1-4\frac{\sigma^2}{\rec^{2}}+\ldots\right) & \rec \gg
\sigma
\end{cases}
\end{equation}
with $c$ a coefficient\footnote{Note that in the limit of very small $\ds$, $\ds
< \exp(-c\rec^2/\sigma^2)$, the expressions break down. This is unlikely to be
relevant in practice.}. At small $\rec$, the fixation probability decreases very
rapidly with decreasing $\rec$. This stems from the fact that mutations in
individuals from the high fitness tail of the Gaussian fitness distribution have an exponentially
greater chance of fixing than those in the bulk. At large $\rec$, by contrast, the genetic
background on which the mutation arises plays only a minor role, since the rate
of switching background is larger than the selection differentials. While
starting out on a fit background gives a mutation a slight advantage, mutations
on any background have a significant chance of fixing. For large $r$, the result for $\pfix$ is
therefore given by small perturbations of the result without background
interference: $\pfix\approx \ds$.

The expressions for $\pfix$ presented above depend on the variance in fitness
$\sigma^2$. In an evolving population  the variance is not a free parameter. When
the effects of mutation on the {\it  bulk} of the fitness distribution can be
neglected, as they can here, the variance is equal to the rate of adaptation,
$v$. The rate of adaptation, in turn, is given by product of the rate at which
beneficial mutations enter the population $N\mut$, the magnitude of their effect
$\ds$ and their probability of fixation.
\begin{equation}
	v=N\mut \ds \sigma\psfix(\rec/\sigma, \ds/\sigma)=\sigma^2
\end{equation}
The rate of adaptation, $v$, can therefore be obtained by solving self-consistently for 
$\sigma$ in the above equations. Substituting our result for $\pfix$ and
ignoring  logarithmic factors in the arguments of large logarithms, we find,  for the free
recombination model,
\begin{equation}
v\approx 
\begin{cases}
\vspace{0.2cm}
2\ds^2\left(\frac{\rec}{\ds}\right)^2 \frac{\log N\mut}{\log^2 \rec/\ds} &
1\ll\frac{\rec^2}{\ds^2}\ll N\mut/\log{N\mut}\\ 
N\mut \ds^2\left(1-\frac{4N\mut \ds^2}{\rec^{2}}+\ldots\right) &
\frac{\rec^2}{\ds^2}\gg 4N\mut
\end{cases}
\end{equation}
Contrary to intuition, $v$ is proportional to $\log N\mut$ rather than $N\mut$
both for low $\rec$ at fixed $N\mut\gg 1$,  and at fixed $\rec$ for sufficiently  large populations sizes, $N$. This indicates that 
the interplay between mutations --- especially their collective effects on fluctuations --- is limiting the
rate of adaptation \cite{Gillespie:2001p9636}.  
As in the asexual case, because of interference
between mutations, only a small fraction  $\sim \log(N\mut)/N\mut$ of the beneficial
mutations fix --- the rest are wasted. However, this fraction increases with
increasing rate of recombination leading to $v$ increasing as $\sim\rec^2 \log
N\mut$, until it saturates at $N\mut \ds^2$, which is the limit of
independently fixating mutations. In this high recombination limit, the rate of
adaptation is limited simply by the supply of beneficial mutations
$N\mut$. Very similar results for the dependence of $v$ on $r$ and $N$ are
obtained for the communal recombination model, differing only by coefficients
inside logarithms and by correction terms.

In the {\it minimal recombination model}, for which only
one locus is exchanged at a time, the behavior is slightly different. For the fixation probability, we find
\begin{equation}
\label{eq:minimalrec_establishmentprob}
\pfix \sim e^{-\sigma^2/2\rec^2 + \ds\sigma^2/\rec^3} \ .
\end{equation}
In contrast to the other models for which recombination results in a macroscopic
change of the genotype, the minimal recombination model  only changes one locus 
at a time.  This results in a slightly  weaker
dependence of $\pfix$ on the recombination rate for $\rec\gg \ds$.
Self-consisting the fitness variance as before determines the speed of
adaptation to be
\be 
v\approx 2\rec^2\log(N\mut) (1+2\ds/\rec) \ . 
\ee
Surprisingly, this result in essentially independent of $\ds$ for $\rec\gg \ds$: 
the larger increase in the fitness per sweep is almost perfectly 
canceled by the decrease in establishment probability.  Note that this 
model is defined with recombination rate $\rec$ per locus so that the 
total number of recombinations in time $1/\rec$ is far more than in the 
other models. But the time for turnover of the genome and loss of linkage 
is of order $1/\rec$ and thus $\rec$ is the useful quantity to compare with the other models.  

\subsection{Simulations}
\begin{figure}[htp]
\begin{center}
\includegraphics[width=0.9\columnwidth]{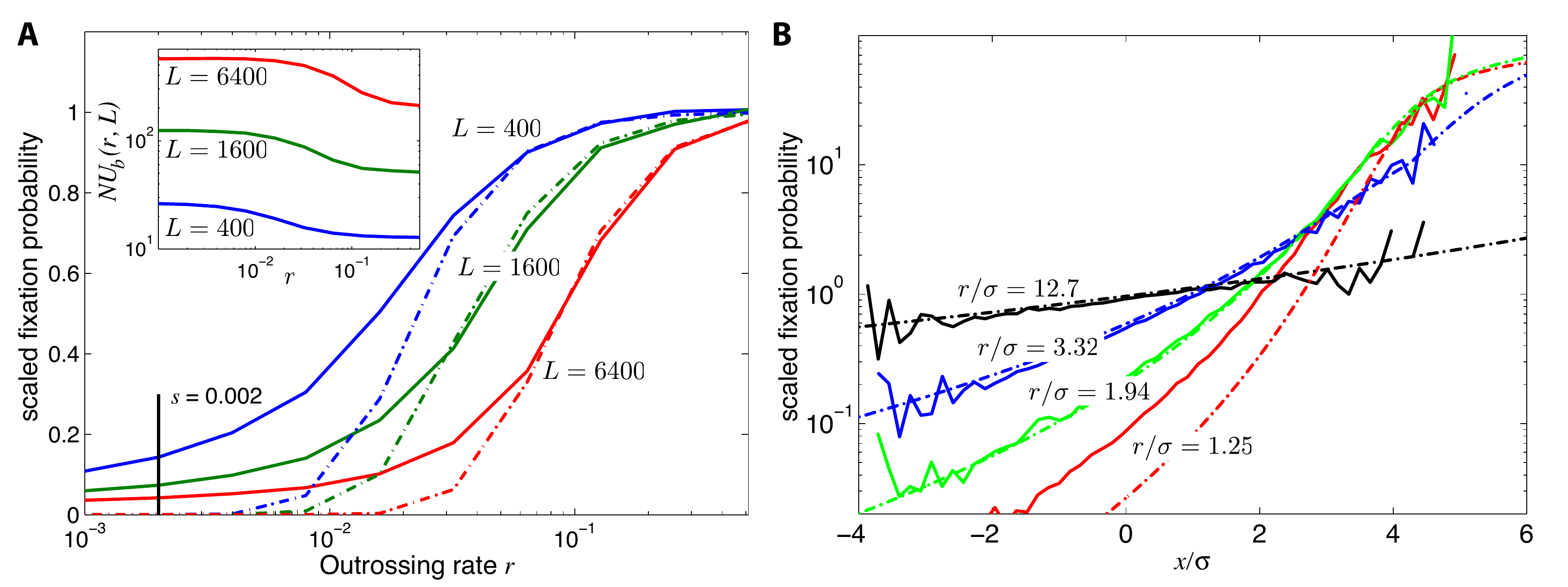}
  \caption{Fixation probabilities in recombining populations. Panel A shows 
the mean fixation probability normalized to the value in the high 
recombination limit as a function of $\rec$ for three different genome sizes
$L$ (with $\ds=0.002$, $N=20000$). The effective rate of beneficial mutations
$N\mut$ is shown in the inset (see main text). The scaled fixation 
probability in the simulation (solid lines) is calculated as $v/2N\mut \ds^2$
and compared to the analytic results for the scaled establishment probability
$\pfix(\rec,\sigma)/\ds$ (dashed lines). The latter are obtained through 
numerical solution of Eq.~(\ref{eq:fixprob_rescaled}) using
$\sigma^2$ observed in simulations. The agreement between simulations and the
analytic approximation improves with increasing $L$, i.e.~increasing $N\mut$,
as expected. Panel B: The scaled fixation probability as a
function of the rescaled background fitness $\x/\sigma$ (relative to the mean). 
The solid lines are simulation results for $\w(\x)$ divided by $2\ds$ using
$L=6400$ and $\rec=0.512,\ 0.128,\ 0.064$ and 0.032: the corresponding
values of the key ratio $\rec/\sigma$, which determines the shape of $\w(\x)$, are
indicated in the figure. The dashed lines are predictions for $\w(\x)/\ds$ 
obtained via numerical solutions of Eq.~(\ref{eq:fixprob_rescaled}). Note that the 
simulation data becomes noisy when the frequency of $\x$ in the population is
around $1/N$.}
  \label{fig:pfix}
\end{center}
\end{figure}

In writing down Eq.~(\ref{eq:fixprob_rescaled})  for the establishment
probability of a beneficial mutation, we have assumed that the distribution of
fitness in the population is gaussian and that correlations and fluctuations are
negligible. Thus it is useful to compare the analytic results to individual-based
simulations of an evolving population. In our simulations, we use a discrete
generation scheme, where each individual produces a Poisson distributed number of
gametes with parameter $\exp(\X-\bar{\X}+\alpha)$. The population size,  $\tilde N$,
is kept approximately constant with an average of $N$ by adjusting the overall rate of replication
through $\alpha=(1-\tilde N/N)\log 2$.  Each individual is represented by a
string of integers, where each bit represents one locus. Recombination,
approximating the free recombination model, is implemented as follows: Each generation, 
gametes are randomly placed  into a pool of asexual gametes with probability $1-\rec$ and into a pool of sexual
gametes with probability $\rec$. The asexual gametes are placed unchanged into
the next generation. The sexual gametes are paired at random and their genes
reassorted to produce haploid offspring. Whenever one locus becomes monomorphic
--- via fixation or extinction of an allele --- , one individual is chosen at
random and a mutation introduced at that specific locus. This allows us to make
optimal use of the computational resources by keeping as many polymorphic loci as
possible. However, this scheme renders the beneficial mutation rate, $\mut$, a
dependent quantity which, as shown in Fig.~\ref{fig:pfix}, increases with $L$ and
decreases with $\rec$. The effective total rate for new beneficial mutations,
$N\mut$, can be determined simply by measuring the average rate at which the new
mutations are introduced (which, the way the simulations are done, is the sum of
the extinction and fixation rates).

Figure~\ref{fig:pfix} shows the mean establishment probability as a function of
the outcrossing rate $\rec$, for different values of $L$ which is roughly
proportional to $N\mut$ (see above). The establishment probability is small at
small $\rec$ but increases sharply and saturates at high $\rec$ at $\pfix=2\ds$
--- the usual single-locus result. The upturn of $\pfix$ occurs at larger $\rec$
for larger $N\mut$, in accord with the prediction that the high recombination
limit is reached when $\rec$ substantially exceeds $\sigma$. The agreement
between the analytic predictions  in the gaussian {\it Ansatz} (via numerical solution of Equation
\ref{eq:fixprob_rescaled}) and the simulation improves as $N\mut$ increases,
suggesting that, as we expect,  the approximations used become valid for large
populations. Note, however, that the corrections to the asymptotic results are
quite large  as the basic small parameter of the gaussian {\it Ansatz} is inversely proportional to
$\log(N\mut)$. The right panel of Figure~\ref{fig:pfix} shows $\w(\x)$, i.e.~the
establishment probability of a mutation arising on background $\x$, measured in
simulations together with the predictions obtained from numerical solution of
Eq.~(\ref{eq:fixprob_rescaled}). At outcrossing rates much larger than $\sigma$,
the fixation probability increases only slightly with the background fitness and
all new mutations have a substantial  chance --- of order $\ds$ ---  to
establish. With decreasing $\rec/\sigma$, the establishment probability becomes a
steeper function of the background fitness and only those mutations arising on
high fitness backgrounds have a significant chance of establishment. Note that at
$\rec/\sigma\approx 1$, $\w(\x)$ measured in simulations decays less rapidly at
small $\x$ than the solution of Eq.~(\ref{eq:fixprob_rescaled}). These deviations
are probably due to fluctuations of the high fitness edge and the width of the
distribution which are ignored in the analysis.  However, as discussed below,
such fluctuations decrease with increasing $N\mut$ as long as $\rec\gg\ds$.

\section{Analysis of Establishment Probability}
\begin{figure}[htp]
\begin{center}
  \includegraphics[width=0.5\columnwidth]{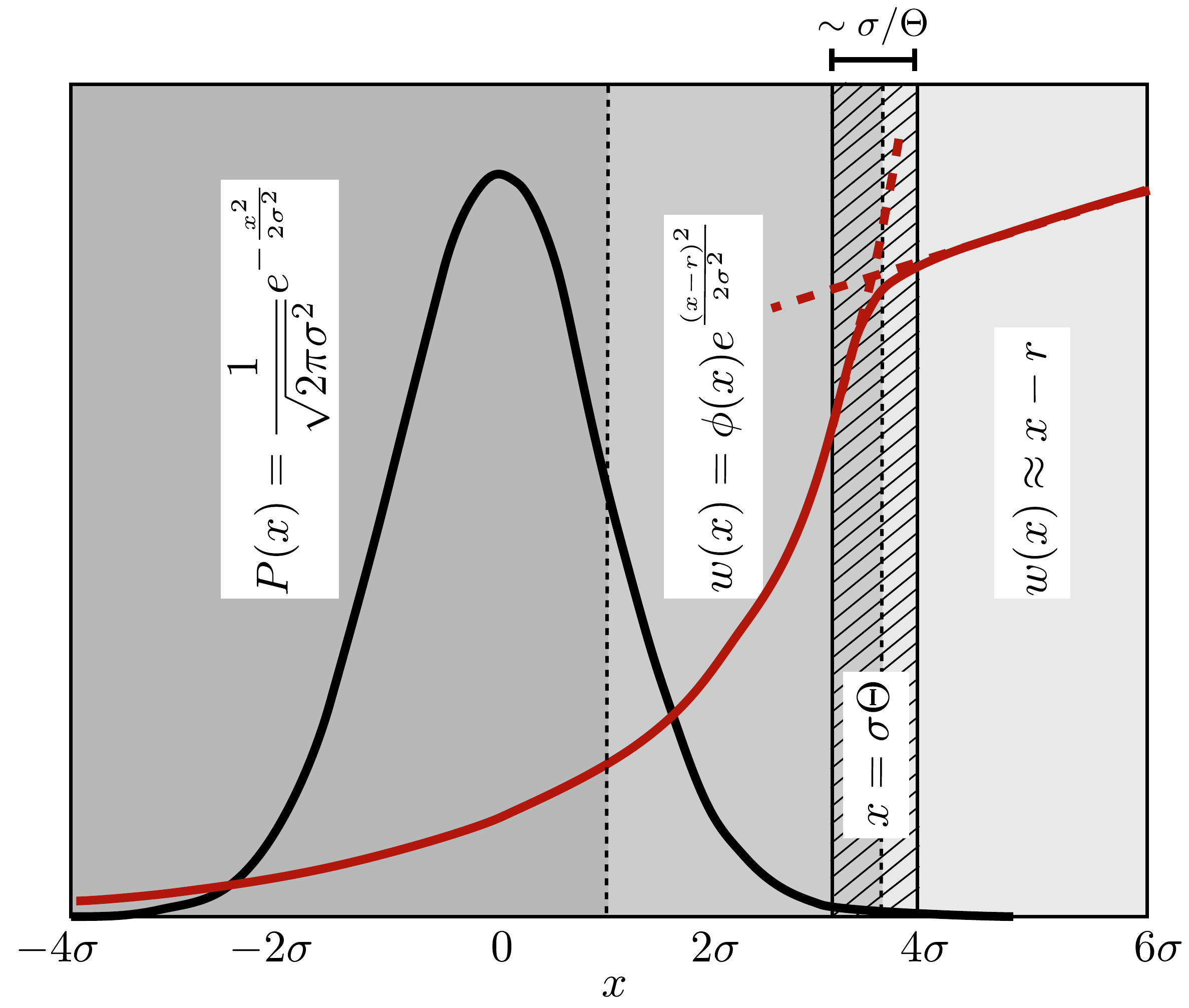}
  \caption[labelInTOC]{Asymptotics of the establishment probability. The 
  fitness distribution $\fitdis(\x)$ of the population is shown in black, a sketch
  of the establishment probability, $w(x)$, is shown in red for $\rec\ll\sigma$. At low
  $\x$, $\w(\x)$ is small and depends sensitively on the recombination model, 
  at intermediate $\sigma <\x< \sigma\B$, $\w(\x)$ increases sharply as $\sim
  e^{\frac{(\x-\rec)}{2\sigma^2}}$, modulated by a slowly varying function
  $\phi(\x)$ which depends on the recombination model. At still larger $\x$, 
  beyond $\sigma\B$, the quadratic term in Eq.~(\ref{eq:fixprob_rescaled}) 
  becomes important, forcing $\w(\x)$ to saturate at $\x-\rec$. The width of
  the crossover region is of the order of $\sigma/\B$.  }
  \label{fig:fixprob_illustration}
\end{center}
\end{figure}

We now turn to a derivation of the results given for the establishment
probability in Eqs.~(\ref{eq:establishmentprob}) and
(\ref{eq:minimalrec_establishmentprob}), which requires solving
Eq.~(\ref{eq:fixprob_rescaled}). We first study the case of $\ds\ll \rec\ll \sigma$ applicable, as we shall see, for very large populations. We proceed by
analysing Eq.~(\ref{eq:fixprob_rescaled}) in different regimes of $\x$. At large
positive $\x-\rec\gg \sigma$, the equation reduces to
$(\x-\rec)\w(\x)\approx\w^2(\x)$ with solution $\w_>(\x)\approx \x-\rec$,
as illustrated in figure \ref{fig:fixprob_illustration}. In this regime, $\w(\x)$
is independent of the recombination model and is simply given by the
establishment probability of a mutation in the absence of any gains from
recombination (but with the clonal growth rate reduced by $\rec$ due to recombination). Establishment
is driven by clonal expansion and contributions from recombination are negligible. (But we shall see 
that there are almost no individuals in the population with such high fitness.)
In the opposite regime, at large negative $\x$, $\w(\x)$ is small and the quadratic
term, as well as the perturbation $\ds\w(\x)$ can be neglected. The resulting
linear equation for $\w_<(\x)$ valid for small $\x$ is
\begin{equation}
\label{eq:linear_fixprob}
(v\partial_\x - \x+\rec ) \w_<(\x) - \rec\int d\y K(\x,\y)
\w_<(\y) \ = 0 \ .
\end{equation}
In this regime, the solution depends sensitively
on the recombination model. This is intuitive, since the only --- and very unlikely --- 
way for a mutation at $\x\ll 0$ to fix is to recombine onto better backgrounds.
We will verify below for each model separately, that the crossover from the
linear regime, $\w_<(\x)$, to the saturated behavior at large $\x$, $\w_>(\x)$, 
occurs rather sharply around $\x/\sigma=\B\gg 1$. At intermediate
$\sigma<\x<\sigma\B$, the establishment probability $\w_<(\x)$ increases steeply
(while remaining small enough for the quadratic term to remain negligible). Individuals in this intermediate regime are much fitter than the average individual so that
recombination usually leads to less fit offspring. Hence the
recombination term is of secondary importance in this range and  $\w_<(\x)$ is
governed by the first term in Eq.~(\ref{eq:linear_fixprob}). The solution to
Eq.~(\ref{eq:linear_fixprob}) is therefore of the form $\w_<(\x)=\phi(\x) e^{(\x-\rec)^2/2\sigma^2}$, where $\phi(\x)$ is a slowly
varying function that depends on the recombination model. This behavior can be
interpreted in terms of the dynamics of a genotype with initial fitness $\x$.
The genotype will expand clonally with rate $\x-\rec$, giving rise
to approximately $n_\x \sim e^{(\x-\rec)t-vt^2/2}$ unrecombined descendants
after $t$ generations. Since each of these could give rise to a lineage which
will fix, in this regime $\w(\x)$ is proportional to $\int n_\x(t) dt$, which 
increases rapidly with $\x$. This is valid up to just below the crossover where
the quadratic term, $\w(\x)^2$, starts to be important, see
fig.~\ref{fig:fixprob_illustration}.

Note that the amplitude of $\w_<(\x)$ is left undetermined by the homogeneous
linear equation (\ref{eq:linear_fixprob}) and hence the location $\B$ of the
crossover is not fixed. To insure that $\w(\x)$ solves the complete
Eq.~(\ref{eq:fixprob_rescaled}), we need to impose the ``solvability
condition'' Eq.~(\ref{eq:solv_cond}) as an additional constraint. The solvability condition involves the first and second
moment of $\w(\x)$ with respect to the fitness distribution $\fitdis(\x)$. The
first moment is dominated by  small and intermediate $\x$
since $\fitdis(\x)\w(\x)$ decreases with $\x$. The second moment, however, is
dominated by a narrow range of width $\sim\sigma/\B\ll \sigma$ around the 
crossover point $\sigma\B$:  for $x\approx \sigma\B$, $\fitdis(\x)\w_<(\x)^2$ increases rapidly with $\x$, while
$\fitdis(\x)\w_>(\x)^2$ decreases rapidly. 
The ``solvability condition'' (\ref{eq:solv_cond}) then becomes
\begin{equation}
\label{eq:simple_solv_cond}
\ds\pfix\approx \frac{\sigma\B}{\stp} e^{-\B^2/2}
\end{equation}
giving us a relation between $\pfix$ and $\B$. 
To analyze the behavior of the various models it is convenient to rescale the
rates and  fixation probabilities as
\be 
\label{eq:rescaling}
\xs=\x/\sigma,  \ \ \ \rs=\rec/\sigma, \ \ \ \dss=\ds/\sigma, \ \ {\rm and} \ \
\ws(\xs)=\w(\xs\sigma)/\sigma \ .
\ee
Utilizing the transform,  
\be 
\OM(z)\equiv \int_{-\infty}^\infty
\frac{d\xs}{\stp}e^{-\frac{(\xs-z)^2}{2}}\ws(\xs) \ ,
\ee 
turns out to be informative: note that the scaled fixation probability is  $\psfix\equiv\pfix/\sigma=\OM(z=0)$. By
integrating the rescaled Eq.~(\ref{eq:fixprob_rescaled}) over the kernel
$\frac{1}{\stp}e^{-\frac{(\xs-z)^2}{2}}$, we obtain an equation for $\OM(z)$ of the
form 
\begin{equation}
{\cal L}\OM =\int_{-\infty}^\infty
\frac{d\xs}{\stp}e^{-\frac{(\xs-z)^2}{2}}\ {\cal J} \ \ws(\xs) =
\dss\OM(z) - \int_{-\infty}^\infty
\frac{d\xs}{\stp}e^{-\frac{(\xs-z)^2}{2}}\ws(\xs)^2\,
\end{equation}
which defines  for each model a linear operator $\cal L$ acting on $\OM(z)$
(${\cal J}$ is the linear operator defined by the left hand side of Eq.~\ref{eq:fixprob_rescaled})).
The integral over $\ws(\xs)^2$ is again dominated by the crossover region and
can be evaluated using $\ws(\B)\approx \B$ and the (scaled) crossover width $\sim\B^{-1}$ 
\be
	{\cal L}\OM \approx\dss\OM(z) -\frac{\B}{\stp} e^{-(\B-z)^2/2}= \dss\OM(z)
	-\dss\OM(0) e^{\B z-z^2/2} .
\ee
The last step was obtained by substituting Eq.~(\ref{eq:simple_solv_cond}). 
The condition that the solution $\ws_<(\xs)$ joins smoothly to the 
saturated solution $\ws_>(\xs)$ and hence only grows slowly for large $\xs$,  
translates into the condition that $\OM(z)$ does not diverge at any fixed $z$: 
it should be an analytic function of $z$. We now examine separately the
different models, simplest first.

\subsection{Communal recombination model.}
In the communal recombination model, the genotypes of offspring are independent
of their parental fitness, which makes this model particularly simple. It can,
in fact, be solved exactly, as shown in Appendix \ref{sec:AppCommunal}, or, in the regimes of interest, by matched asymptotic expansions. But it is more  instructive to  proceed 
with the approximate but more general and asymptotically exact analysis 
outlined above. The equation for $\OM(z)$ reads 
\be
{\cal L_C}\OM \equiv(\rs-z)\OM(z)-\rs \psfix=\dss\OM(z)-\dss\OM(0)
e^{z\B-z^2/2}\, 
\ee
which can be solved trivially. But in general it has a pole at $z=\rs-\dss$. This
pole has to be canceled, since we know that $\ws(\xs)$ saturates at $\xs=\B$ and
$\OM(z)$ cannot develop a singularity. Hence, we must have
$e^{\B(\rs-\dss)-(\rs-\dss)^2/2}=\rs/\dss$ to eliminate the pole. Solving for
$\B$ and substituting it into the solvability condition (\ref{eq:simple_solv_cond}) yields
\be
\psfix \approx \
\frac{\log(\rs/\dss)}{\dss(\rs-\dss)\stp}e^{-\frac{1}{2(\rs-\dss)^2}\left(\log(\rs/\dss)+\frac{(\rs-\dss)^2}{2}\right)^2}
\approx \frac{\log(\rs/\dss)}{\dss^{1/2}\rs^{3/2}\stp}e^{-{\log ^
2(\rs/\dss) \over 2 \rs^2 }} \ .
\ee
The last approximate equality is correct to  leading order in $\dss / \rs \ll 1$.

\subsection{Free recombination model}
In the free recombination model, the offspring obtains on average half of its
genome from either parent. The parent carrying the new allele mates with a random
member of the population: thus after recombination the average fitness of the genotype carrying
the new allele is half as far from  the population mean fitness as it was before recombination. As a
result of this correlation between parents and offspring, the operator ${\cal
L_I}$ for the free recombination model is more complicated and couples $\OM(z)$
to $\OM(z/2)$.  
\be
\label{eq:freerecOmega}
{\cal L_I}\OM \equiv  (\rs-z)\OM(z) - \rs\OM(z/2) \approx \dss\OM(z)-\dss
\OM(0) e^{z\B-z^2/2} 
\ee 
where, as before, $\psfix = \OM(0)$. Neglecting the $e^{-z^2/2}\approx 1$ on
the right hand side (we need only consider $z\ll 1$ since $\rs\ll 1$), we can
analyze this as a power series in $z$ writing $\OM(z)=\sum_n \OM_n z^n$ finding 
\be
\label{eq:hermite_coeff}
\frac{\OM_n}{\OM_0}= \prod_{k=1}^n\frac{1}{\rs-\dss-\rs2^{-k}} -\dss
\sum_{j=1}^n\frac{\B^j}{j!}\prod_{k=j}^n\frac{1}{\rs-\dss-\rs2^{-k}} \ . 
\ee 
As the first part would yield ratios of successive terms which approach
$1/(\rs-\dss)$ for large $n$ and again induce a pole at $z=\rs-\dss$, this has to
be canceled by the second inhomogeneous term. The condition for convergence (up
to well beyond the ``almost-pole'' at $\rs-\dss$) is that $\OM_n (\rs-\dss)^n \to
0$ for $n\to \infty$ which requires that \be
1=\dss\sum_{j=1}^\infty\frac{\B^j}{j!}\prod_{k=1}^{j-1}(\rs-\dss-\rs2^{-k})
\approx  e^{\B(\rs-\dss)}\frac{\dss}{\rs}\prod_{k=1}^{\infty}(1-2^{-k}). \ee The
last approximate equality is accurate when  $\dss\ll \rs$ and hence
$\B(\rs-\dss)\gg 1$. Thus we must have \be
\label{eq:XOfreerec}
\B\approx\frac{\log(c\rs/\dss)}{\rs-\dss} \ee with the order-unity coefficient
$c=1/\prod_{k=1}^\infty(1-2^{-k})$. We thus obtain $\psfix$ very similar to the
communal recombination model, \be \psfix \approx
\frac{\log(c\rs/\dss)}{\dss(\rs-\dss)\stp}e^{-\log^2(c\rs/\dss)/2(\rs-\dss)^2}
\ . \ee Note that $\OM(z)$ is approximately the Laplace transform of
$\phi(\xs)=\ws(\xs)e^{-\xs^2/2}$, which can be analyzed perturbatively for small
$\rs$, see Appendix \ref{sec:Appfreerecmodel}. This expansion in $\rs$ reveals
the most probable --- least unlikely --- path of a mutation on a typical initial background to
successively better backgrounds and establishment.

\subsection{Minimal recombination model}
The minimal recombination model can be analyzed similarly: ${\cal L}_T$ is
now a differential operator, and we have  
\be 
{\cal L_T}\OM \equiv (\rs-z)\OM - \rs\psfix +\rs z\frac{d\OM}{dz}\approx
\dss\OM- \dss \OM(0) e^{z\B} \ .
\ee 
This can be explicitly integrated and the behavior for $1\gg z>{\cal O}(\rs)$
found to involve linear combinations of $e^{z/\rs}$ and $e^{z\B}$.  For $\dss\ll\rs$, the condition that
the solution matches correctly onto the non-linearly saturated form for
$\xs\approx \B$, can be shown to be that these two exponentials are almost the
same.  This yields the condition $\B\approx 1/\rs$. In contrast to the other models, $\dss$
only gives corrections to  $\B$.  The fixation probability is then  found  to
be 
\be 
\pfix \sim e^{-1/2\rs^2 + \dss/\rs^3} 
\ee 
which yields a different form for the speed of evolution: 
\be 
v\approx 2\rs^2\log(N\mu) (1+2\dss/\rs) \ . 
\ee

\subsection{High recombination rates}
In the limit of high recombination rate, the crossover to the saturated solution
$\ws_<(\xs)$ occurs far out in the ``nose"  (high fitness tail) of the
population distribution --- further out than any individuals are likely to be. In this regimes, the
assumption that $\int d\xs e^{-\xs^2/2} \ws(\xs)^2$ is dominated by the crossover
region is no longer justified.

To analyse this high $\rec$ regime, we can make use of the expansion of
$\OM(z)=\sum_n z^n\OM_n$, which is equivalent to expanding $\ws(\xs)$ in Hermite
polynomials $\ws(\xs)=\sum_n \OM_n  H_n(\xs)$, where the $H_n(\xs)=(-1)^n
e^{\xs^2/2}\partial_\xs^ne^{-\xs^2/2}$. In the limit of $\rs\gg\dss$, the second
term in Eq.~(\ref{eq:hermite_coeff}) can be neglected for the first few
coefficients and we have $\OM_n/\OM_0=\prod_{l=1}^n \frac{1}{\rs(1-2^{-l})}$ (for
the communal recombination model we have $\OM_n/\OM_0=\rs^{-n}$). The value of
$\OM_0=\psfix$ has to be determined by the solvability condition $\dss\psfix=\int
d\xs/\stp e^{-x^2/2}\ws(\xs)^2$. From the orthogonality of the Hermite
polynomials one finds that
 the right hand side is  simply  $\sum_n n!\OM_n^2$.
Hence, we find for the fixation probability the formal expression
\begin{equation}
\psfix=\OM_0=\dss \left(1+\sum_{n=1} n!\prod_{l=1}^n
\frac{1}{\rs^2(1-2^{-l})^2}\right)^{-1}
\end{equation}
The $n!$  would cause the sum to diverge if it extended to
infinity. But for large $\rs$, this is a valid asymptotic series,
which can be truncated at any finite
number of terms. To zeroth order, one finds in both models $\pfix=\sigma\dss=\ds$
which is simply  the result in a homogeneous population. Including the first two
non-trivial correction terms, one finds
\begin{eqnarray}
\pfix =&\ds\left(1-4\rs^{-2}+\frac{16}{9}\rs^{-4}+\cdots\right) \quad
&\mathrm{free~recombination~model}\\
\pfix =&\ds\left(1-\rs^{-2}-\rs^{-4}+\cdots\right) \quad
&\mathrm{communal~recombination~model}\nonumber
\end{eqnarray}
[Note that the divergence of the expansion for large $n$, for which this approach
breaks down,  is related to the singular dependence of $\psfix$ on $1/\rs$ for
small $\rs$ discussed above.] For the minimal recombination model, the behavior
for large $\rec$ is similar and the expansion in inverse powers of $\rs$ can be
analyzed: we do not carry this out here.

\subsection{Range of validity of analysis}
Throughout the analysis, we have assumed that the fitness distribution of
individuals in the population, $\fitdis(x=\X-\bar{\X}(t))$,  is gaussian, and also that of 
recombinant offspring.  Crucially, for the analysis, we assumed that it remains gaussian  in the high-fitness nose of the
distribution all the way to the crossover point $\B$ which controls the
establishment probabilities.  We  need to justify this {\it Ansatz}.   First, as
noted earlier, we observe that a gaussian  fitness distribution is the exact
traveling-wave solution to the linear recombination model in the absence of
fluctuations: the gaussian approximation should thus be valid throughout the
bulk of the distribution in the limit of very large populations.  Second,  in the absence
of fluctuations (or epistatic interactions which we are ignoring in any case) the
frequencies of alleles at different loci are independent. And third,  if the
establishment probabilities of different beneficial mutations are independent,
then it can be shown that the resulting Poisson process of the establishments together with
random combining of the  alleles with their corresponding frequencies leads to a
distribution of fitnesses whose logarithm averaged over the establishment times,
$\langle \log(\fitdis(\x))\rangle$, is exactly parabolic --- corresponding to a
gaussian distribution.
However, due to fluctuations and correlations, the distribution of fitnesses
will be neither exactly gaussian nor exactly time-independent and
we must check that the non-fluctuating gaussian is a good enough approximation far enough out in the nose in the
large $N$ regimes of interest.

We first check that the sampling of the distribution due to the
finite population size is sufficient.  A population of size $N$ samples a
close-to-gaussian distribution only out to about $\sigma\sqrt{2\log N}$ ahead of
the mean. But this implies that, with the fitnesses of individuals  only weakly
correlated, the crossover region near $\B$ is indeed well sampled by the
population since
\begin{equation}
\B\approx\sigma\frac{\log c\rec/\ds}{\rec}=\sqrt{2\log N\mut}< \sqrt{2\log N} \
.
\end{equation}
The last inequality is valid when the rate of beneficial mutations per genome per
generation, $\mut$, is small as is surely always the case: there are then of
order $1/\mut$ individuals in the population with fitnesses in the crucial
crossover region of the establishment probabilities. Furthermore,  the Gaussian
shape of the fitness distribution will be a good approximation when the
number of polymorphic loci that contribute substantially to the fitness variance
is large. However, the total number of established polymorphic loci
is dominated by low frequency alleles. (The total number of polymorphic loci is
much higher still, but almost all of these are not established and destined to
soon go extinct.) Nevertheless, there are sufficiently many polymorphic sites
with high enough frequencies that they contribute substantially to the fitness distribution. Since sweeps occur at rate $v/\ds$ and since a
sweeping allele is at intermediate frequencies for a few times $1/\ds$
generations, the number of loci, $K$, contributing substantially to the variance is of order
$v/\ds^2 \sim (\rec/\ds)^2\log(N\mut)$. For $\rec\gg s$ these $K$ loci are
approximately in linkage equilibrium, giving rise to a gaussian fitness
distribution with corrections to parabolic $\log(\fitdis(x))$ of order 
$(x/\sigma)^2/K$. At the
crossover point, $\sigma\B$, it can then be checked that the corrections to $\fitdis(\x)$
are small as long as $\rec\gg\ds\sqrt{\log N\mut}$. We  thus expect that this is the condition for validity of the gaussian {\it Ansatz} from which our analytic predictions are obtained.  A more detailed analysis of
the effects of fluctuations, in particular in the crucial ``nose'' of the 
distribution, is left for future work.

\section{Discussion}
We have analyzed in several simple models the dependence of the speed of adaptation  on
the rate of recombination and the population size, focusing on the particularly
interesting behavior in the wide
range of outcrossing rates $ \ds\sqrt{\log N\mut}\ll \rec  < \ds \sqrt{ N\mut/ \log
N\mut }$, or equivalently, on population sizes $N\mut \gg
\frac{\rec^2}{\ds^2}\log(\rec/\ds)$.  In the  high recombination limit and moderate $N$ the conventional analysis of independent fixations holds
and the rate of adaptation (and concomitantly the variance of fitness) are proportional to the total production 
rate of beneficial mutations, $N\mut$.  In contrast, for large populations (with
recombination rates in the intermediate regime) we find adaptation rate $v \sim
\rec^2 \log N\mut$. This change from linear to logarithmic dependence on $N\mut$
indicates that the rate of adaptation is limited by  interference among
multiple simultaneously segregating beneficial mutations rather than by the
supply of beneficial mutations.   This reduction in the rate of adaptation due to
linkage is, qualitatively, the Hill-Robertson effect \cite{Hill_GenetRes_1966}.
Most interestingly, while logarithmic in population size, the rate
of adaptation increases with the rate of recombination as $\rec^2$. Hence our  results
confirm the heuristic arguments by \citeauthor{Fisher_1930} and
\citeauthor{Muller_AmericanNaturalist_1932} and provide a quantitative
framework for identifying conditions favoring sexual reproduction
\cite{Barton_Science_1998, Rice:2002p28396}.

The rate of adaptation is determined by the dynamics of the linkage between new
beneficial alleles and the spectrum of fitnesses of the rest of the genome. This
results in most new mutations being eliminated by their linkage to modestly fit
genomes which rapidly lose out with respect to the steadily increasing average
fitness driven by the anomalously fit genomes. Only those alleles that
either arise on very fit genomes or are lucky enough to recombine to make a very
fit genome will survive long enough for their frequency to grow deterministically
and sweep through the population.  The logarithmic dependence on population size
is similar to that found for purely asexual evolution when multiple beneficial
mutations are present in the population \citep{Desai_Genetics_2007}.  But with
$\rec>\ds$, recombination speeds up the adaptation by allowing new mutations
that arise on modestly fit backgrounds to recombine to very fit backgrounds and
thereby fix.

We have shown that the typical number of simultaneously segregating alleles at
intermediate frequencies is on the order of $K\sim \rec^2/\ds^2\log N\mut$. For
$\rec\gg \ds$, the number of possible combinations of these sweeping loci
therefore dramatically exceeds the population size. This implies that the limit
of ``infinite'' population size, for which {\it each} genotype is  well-sampled
is unattainable at fixed recombination and beneficial mutation rate.  On the
contrary, sampling becomes sparser and the benefits of recombination more
pronounced in larger populations. The population size dependence of the
beneficial effects of recombination has been a subject of considerable
theoretical debate
\citep{Crow_AmericanNaturalist_1965,Smith_AmericanNaturalist_1968,Barton:2005p982}.
The increased advantage of sexual reproduction in large population  has been
demonstrated in model simulations by \citet{Iles:2003p28305}. It has also been
observed experimentally by \citet{Colegrave:2002p6888}, who studied this
phenomenon in an evolution experiment with \beast{C.~reinhardtii}.

\subsection{Relationship to other recent work}
The description of the spread of beneficial alleles in space as a traveling wave
goes back to \citet{Fisher_1930}. The notion that adaptation of a panmictic
population can be described as a travelling wave in  {\it fitness} was introduced by
\citet{Kepler:1995p26819} and \citet{Tsimring:1996p19688}. In these effectively
deterministic models, the velocity of the pulse is determined by the size of the
population through a modification of the deterministic solution at the high
fitness edge --- the ``nose" or ``front" --- to approximate the crucial stochastic behavior
near the nose \cite{Brunet:1997p18870}. These concepts were applied to
recombining populations by
\citet{Rouzine_Genetics_2005} and \citet{GheorghiuSvirschevski:2007p17402}
who studied the rate of (transient) adaptation when selection acts on standing
variation.
\citet{Cohen_PhysRevLett_2005,Cohen:2006p5005}
studied continuing evolution with a large supply of beneficial mutations
available in a model that is related to our ``minimal recombination" model.  Both
approaches focused on the overall distribution of fitnesses within the population
and  the primary role of recombination they considered was to maintain a near
gaussian shape of the fitness distribution, achieved by producing higher fitness
individuals and thereby advancing the  nose.  Some of the results of the
approximate analytic treatments are related to ours, including the $\log N$
scaling  of the adaptation speed  in certain regimes. Yet the actual underlying dynamics
implicit in the approximations used %in \citep{Cohen_PhysRevLett_2005} is very
are very different from what we find here and so is the dependence on parameters.

The key feature of the adaptation with substantial rates of recombination is the
stochastic dynamics of new mutations. The  probability that a new beneficial
mutation  will  sweep to fixation is  determined by its establishment
probability: the probability that it escapes stochastic extinction. The establishment probability depends very strongly on the
distribution of fitnesses of the genetic backgrounds with which the new mutation
can be linked. As the distribution of fitness depends on the velocity, the
steady-state velocity must be  determined by  matching the rate of establishment
of new alleles with the velocity of the deterministic traveling wave describing
the fitness distribution in the population. The latter is driven by the
continuous incorporation of a large number of new sweeping alleles that have
successfully  established at earlier times. At any time there is  thus a broad
distribution of frequencies of the beneficial  alleles.
The primary problem with the earlier analysis is that the distribution and
dynamics of individual allele frequencies is not treated directly and the
approximations implicitly made for their forms  are not consistent with the basic
processes.

In contrast with the asexual traveling wave for which a description in terms
of a simple traveling wave is valid \citep{Desai_Genetics_2007,Rouzine:2008p20864} and the
diversity within the population can be ignored, with any amount of
recombination, the diversity and distribution of allele frequencies is
absolutely crucial. It matters a great deal whether the advance of the fitness
wave occurs via small amounts of each of several new alleles, or all from a
single allele. This information is lost by treatments in terms of the fitness
distribution alone.   Note that  in general this is also true for adaptation
from standing variation: beneficial alleles initially at low frequencies can be
driven extinct by their linkage to different backgrounds. If all are initially
at sufficiently high frequencies to avoid this fate, then neither linkage nor
recombination play much role in the dynamics of the adaptation.

The models we have studied were inspired by facultatively mating organisms, in
which outcrossing occurs at rate $\rec$. \citeauthor{Barton_2009}
(pers.~comm.)~have recently performed a related analysis for obligate sexual
reproduction. In addition to a model with a linear genetic map (see below), they
study the free and minimal recombination models, for which they find similar logarithmic
dependence on the population size and mutation rate. Their discrete generation
models with obligate mating do not reveal the dependence of the rate of
adaptation on the outcrossing rate, one of the results of our analysis, but a
similar behavior is implicit in their results.

\subsection{Extensions and open questions}
In this paper we focused on the effect of recombination with $r>s$ in simple models of mating
without chromosomal organization and without epistasis.  We conclude by
considering going beyond these simplifying limits.

We first consider decreasing the recombination rate. In comparing our analytic
results on the free recombination model with the direct simulations we found good
agreement at high recombination rates which confirms the accuracy of the
simplifying assumptions made in analyzing the model (i.e.~
Eq.~(\ref{eq:fixprob_rescaled})). At lower recombination rates we observed that
our ``mean-field" treatment of the recombination  underestimates the rate of
adaptation. This is due to the gradual appearance of ``fat tails" in the
distribution of fitness: specifically, the high fitness nose of the distribution
decays more slowly than the gaussian assumed in the analysis. The fluctuations in the time of establishment of the
currently intermediate frequency alleles becomes important.  Some of the causes of this can be studied analytically. The primary
effect is the smaller number of segregating loci  --- of order $v/\ds^2 \sim
\rec^2/\ds^2$ --- at low recombination rates.  As the ratio $\rec/\ds$ decreases
further,  the acquisition of further beneficial mutations near the nose of the
distribution --- which dominates the asexual evolution --- starts to become
important. Correlations between loci caused by this process and other  sources,
will also play important roles.

The behavior of the leading edge of the fitness distribution is  known to be the
key factor in determining the speed of adaptation in the asexual limit of $r
\rightarrow 0$ \citep{Desai_Genetics_2007} and it will be
of critical importance in the $\rec\ll \ds$ regime. A correct treatment of this
regime, connecting with the known results for asexual adaptation
\citep{Desai_Genetics_2007,Rouzine:2008p20864,Brunet:2008p12980},  requires
analyzing the diversity that is generated by the asexual process and the effects
of small amounts of recombination on this.   It is worth noting that within our
approximations, for the low recombination regime with $\rec\ll \ds$, the
branching process analysis yields an adaptation speed for all three models of the
form $v\sim \ds^2\log(N\mu)/\log^2(\ds/\rec)$ which is a similar form to the
asexual result, $v\approx 2\ds^2\log(N\sqrt{\mu\ds})/\log^2(\ds/\mu)$. This
suggests that in spite of the breakdown of the assumptions, the approximations
may give reasonable results, although not asymptotically accurate ones,  even for
$\ds\gg\rec\gg\mu$.  But we leave this regime, which is particularly important
for microbes with rare genetic exchange, for future investigations.

Our analysis has focused on the simple approximation of additive growth rate
(equivalent to multiplicative fitnesses in a discrete-generation model). 
Some of the most interesting extensions of the present models would include epistasis ---
i.e. genetic interactions --- which makes the effect of each allele  explicitly
dependent on its genetic background. This dependence can be very complex
resulting in low heritability of fitness, in the sense that the fitness of
recombinant progeny may be only weakly correlated with the fitness of the
parents. Remarkably, in the limit of very strong epistasis
\citep{Neher:2009p22302} the establishment probability of an allele is described
by a model which reduces to the communal recombination model described above. The
speed of adaptation is, however, determined by a different self-consistency
condition which will be presented elsewhere.  In general, how to setup --- never
mind analyze! --- instructive models of evolutionary dynamics with epistasis
between  many segregating loci, is largely an open field.

Another important simplification in the free recombination model studied here is
the random reassortment of the parental alleles ignoring the
physical arrangement of the genes.  More realistic models would account for
the linear arrangement of genes on the chromosomes such that chromosomal
proximity implies low recombination rate. In this case, the number of
independently transmitted loci in the event of mating is the product of the
number of chromosomes and the crossovers per chromosome. When the number of
substantially polymorphic loci  is sufficiently large, the free recombination
approximation will certainly break down. But in facultatively mating organisms
where periods of asexual reproduction are interspersed by outcrossing events much
reassortment can occur. Indeed, some facultative outcrossers have high crossover
rates (e.g. \beast{S.cerevisiae} \cite{Mancera:2008p14765}). In this case  the
free recombination model can have a reasonable regime of validity.  More
generally, the fact that our three rather different models yield similar behavior
for the adaptation rates at large population sizes suggests that the forms of the
dependence on parameters --- especially speed proportional to $\log(N\mut)$ ---
may be valid much more broadly.  Arguments to be
presented elsewhere suggest that the balance between the  lengths of linked
regions and the number of polymorphic loci in them can result in $v\sim \rec\ds
\log(N\mut)$ in some regimes.  Significant progress in the analysis of the rate of
adaptation with linear chromosomes has recently been made by
\citeauthor{Barton_2009}. They invoke a scaling argument and use a perturbative
analysis of nearby pairs of segregating loci to derive an expression for the rate
of adaptation. In this approximation, the rate of acquisition of beneficial
mutations tends to an upper limit independent of the population size, selection
coefficient, or mutation rate, being solely determined by the map length: in our
notation this would be equivalent to $v \approx C\rec\ds$ with $C$ a constant.
Note that this is similar to the conjecture quoted above but without the
$\log(N\mut)$ factor. To check whether the approximations are accurate with many
concurrent sweeps it will be necessary to go beyond the perturbative analysis of
\citeauthor{Barton_2009}.  Furthermore, the interplay between the effectively
asexual evolution of short regions of the chromosome that are linked for long
times, and recombination between and within them, needs to be understood and
could well change the behavior qualitatively.

The challenges of understanding evolutionary dynamics in the presence of many
beneficial alleles and recombination between linear chromosomes, and of
understanding the effects of epistatic genetic interactions, provide many
important open problems.

{\bf Acknowledgments: }
We would like to thank Nick Barton for sharing a preprint of his work and
commenting on the manuscript and are grateful to two anonymous referees for
numerous and exceptionally useful suggestions. This research was supported in part by the
National Science Foundation under Grant No. PHY05-51164 (RAN and BIS) and 
the Harvey L.~Karp Discovery Award to RAN.

\appendix
\section{Exact solution of the communal recombination model}
\label{sec:AppCommunal}
In the communal recombination model the genotype of
recombinant offspring is assembled at random from the alleles segregating in the
population and therefore independent of the fitness of the parents. The
equation describing the establishment probability,
Eq.~(\ref{eq:fixprob_rescaled}), therefore simplifies to 
\begin{equation}
\label{eq:apprescaled_fixprob}
\partial_\xs \ws(\xs)=\rs\psfix+(\xs+\dss-\rs)\ws(\xs)-\ws(\x)^2,
\end{equation}
where all rates, the fitness and $\ws(\xs)$ have been rescaled by the standard
deviation of the fitness distribution, as in  Eq.~(\ref{eq:rescaling}). The
quadratic term can be removed by substituting $\ws(\xs)=\frac{\partial_\xs\psi(\xs)}{\psi(\xs)}$, 
which gives rise to the equation
\begin{equation}
\label{eq:psi}
\partial^2_\xs\psi(\xs)-\rs\psfix\psi(\xs)-(\xs+\dss-\rs)\partial_\xs\psi(\xs)=0
\end{equation}
A second substitution of $\psi(\vartheta)=e^{\vartheta^2/4}\phi(\vartheta)$ with
$\vartheta=\xs+\dss-\rs$ maps Eq.~(\ref{eq:psi}) onto the parabolic cylinder
equation
\begin{equation}
\partial^2_y\phi(\vartheta)-\left(\rs\psfix-\frac{1}{2}+\frac{\vartheta^2}{4}\right)\phi(\vartheta)=0
\end{equation} 
The solution with the correct asymptotic behavior is
$\psi_{\rs\psfix}(\xs)=e^{\vartheta^2/4}U(\rs\psfix-1/2,\vartheta)$ and has the integral
representation (\citet{abramowitz}, formula 19.5.1)
\begin{equation}
\psi_{\rs\psfix}(\xs)=\int_0^\infty d\lambda e^{\vartheta
\lambda-\lambda^2/2}\lambda^{\rs\psfix-1}.
\end{equation}
From $\psi_{\rs\psfix}(\xs)$, we obtain $\ws(\xs)$ by taking the log
derivative $\ws_{\rs\psfix}(\xs)=\partial_\xs \log \psi_{\rs\psfix}(\xs)$. The asymptotics
of $\ws_{\rs\psfix}(\xs)$ in the different regimes are
\begin{equation}
\label{eq:piecewise_pool}
\ws_{\rs\psfix}(\xs)=\begin{cases}
     \frac{\rs\psfix}{\rs-\xs-\dss}\left(1-\frac{1+\rs\psfix}{(\rs-\xs-\dss)^2}\right) & \xs\ll\rs-\dss\\ 
     \sqrt{\pi} \rs \psfix e^{(\xs+\dss-\rs)^2/2} & \rs-\dss\ll\xs\ll\sqrt{-2\log \rs \psfix}+\rs-\dss\\ 
     \xs+\dss-\rs & \xs\gg \sqrt{-2\log \rs \psfix}+\rs-\dss \ ,
     \end{cases}
\end{equation}
as found  via the perturbative scheme in the main text.
The fixation probability entered Eq.~(\ref{eq:apprescaled_fixprob}) as a free
parameter and has to be fixed such that $\int
\frac{d\xs}{\stp}e^{-\xs^2/2}\ws_{\rs\psfix}(\xs)=\psfix$, which results in a
very similar condition for $\psfix$ as the solvability condition of the
perturbative scheme used in the main text.

\section{The low recombination limit of the free recombination model}
\label{sec:Appfreerecmodel}
In the intermediate regime where the recombination term and the quadratic term in
Eq.~(\ref{eq:fixprob_rescaled}) are both small, the fixation probability is of
the form $\ws(\xs)=\phi(\xs)e^{(\xs+\dss-\rs)^2/2}$, where $\phi(\xs)$ is a slowly
varying function compared to the gaussian growth term. Ignoring the quadratic
term, the equation for $\phi(\xs)$ reads
\begin{equation}
\partial_\xs\phi(\xs)=2\rs e^{-\xs(\rs-\dss)+\frac{3}{2}(\rs-\dss)^2}\int
\frac{d\ys}{\sqrt{6\pi}}e^{-\frac{(\ys-(2\xs-3(\rs-\dss)))^2}{6}}\phi(\ys)
\end{equation}
Hence, the dominant contribution to the recombination term comes from
$\ys=2\xs-3(\rs-\dss)\approx 2\xs$. The function $\phi(\xs)$, however, drops to
zero rapidly beyond $\B$, implying $\phi(\xs)$ constant in the interval
$\B/2<\xs<\B$. 

To study the behavior of $\phi(\xs)$ more systematically, it is
useful to rearrange Eq.~(\ref{eq:freerecOmega})
\begin{equation}
\label{eq:iter}
\OM(z)=\frac{\dss\psfix e^{z\B}-\rs\OM(z/2)}{z-\rs}\ ,
\end{equation}
where we assumed $\rs\gg\dss$ and $z\ll 1$ such that $\dss$ in the
denominator and $e^{-z^2/2}$ can be neglected. Assuming small $\rs$, this
equation can be solved iteratively. The two terms on the right, however, have
to be matched to cancel the pole at $z=\rs$, which can be done by adjusting $\B$
for each order in the iterative solution. Starting with $\OM^{(0)}(z)=\psfix$, we have
\begin{equation}
\OM^{(1)}(z)=\frac{\dss\psfix e^{z\B_1}-\rs\psfix}{z-\rs}\ ,
\end{equation}
with $\B_1=\frac{\log \rs/\dss}{\rs}$. Iterating Eq.~(\ref{eq:iter}), it is
found that $\B_k=\frac{\log c_k\rs/\dss}{\rs}$ with $c_k\approx\prod_{n=1}^{k-1}
\frac{1}{1-2^{-n}}$, which is rapidly converging to the value of the crossover
point found by power series expansion of $\OM(z)$ in Eq.~(\ref{eq:XOfreerec}).
The solution to $k$-th order reads
\begin{equation}
\label{eq:Laplace_iterative}
\OM^{(k)}(z)=\dss\psfix\sum_{j=0}^{k-1}
\frac{(-r)^{j}e^{z\B_{k-j} 2^{-j}}}{\prod_{n=0}^{j} (z2^{-n}-r)} +
\frac{(-r)^{k}\psfix}{\prod_{n=0}^{k-1} (z2^{-n}-r)}\ ,
\end{equation}
where all poles are canceled by zeros of the numerator.
For small $z$, $\OM(z)$ is related to the Laplace transform of the function
$\phi(\xs)$ in the variable $z-\rs$.  
\begin{equation}
\OM(z)=\int d\xs e^{-\frac{(z-\xs)^2}{2}}e^{\frac{(\xs-\rs)^2}{2}}\phi(\xs)=
e^{-\frac{z^2}{2}+\frac{r^2}{2}}\int d\xs
e^{\xs(z-\rs)}\phi(\xs)
\end{equation}
Since $\phi(\xs)$ is essentially zero for $\xs>\B$ it is useful to
change variables to $\lpvar=\B-\xs$ and consider the Laplace transform on
$\lpvar\in [0,\infty[$:
\begin{equation}
\OM(z)=e^{-\frac{z^2}{2}+\frac{r^2}{2}}\int d \lpvar
e^{(\B-\lpvar)(z-\rs)}\phi(\B-\lpvar)\approx
e^{\B(z-\rs)}\int_0^{\infty} d \lpvar
e^{-\lpvar(z-\rs)}\phi(\B-\lpvar) \ ,
\end{equation}
where we dropped the $z^2$ and $\rs^2$ terms. We can now backtransform
$\OM^{(k)}(z)$ Eq.~(\ref{eq:Laplace_iterative}) into $\xs$-space and obtain an
approximation for $\phi(\xs)$. The inverse transform of terms of the form
$\frac{e^{-s \tau}}{(s+\alpha)^{n+1}}$ is $\frac{(\lpvar-\tau)^{n}}{n!}
e^{-\alpha(\lpvar-\tau)}u(\lpvar-\tau)$, with $u(x)$ being the Heaviside function. The
most important observation is that the delay $\tau=\B(1-2^{-j})$ is different
for the different orders and that higher order terms come in only below a
cut-off set by this delay:
\begin{equation}
\phi_k(\xs) \approx \sum_{j=0}^{k-1} (-r)^j f_j(\lpvar) u(\lpvar+\B 2^{-j}-\B) =
\sum_{j=0}^{k-1} (-r)^j f_j(\B-\xs) u(\B 2^{-j}-\xs) \ .
\end{equation}
Here, $f_j(\lpvar)$ is polynomial in $\lpvar$ multiplied by a slowly varying
exponential $\exp(\rs \lpvar)$ ($\rs\ll 1$). 
This behavior of $\phi(\xs)$ (and $\ws(\xs)$) has a simple interpretation:
For $\B/2^j <\xs <\B/2^{j-1}$ the least unlikely way for a new mutation initially with a 
background fitness $\xs$ to fix is to recombine $j$ times each time 
getting closer to the front at $\B$ beyond which it can rise to a high level
without further recombination. 

\section{Minimal recombination model}
\label{sec:AppMinimalRec}
In the minimal recombination model, the allele at each locus is exchanged
for a random allele from the population at rate $\rec$. Let the locus $i$ of a
particular individual be in state ${\bf s}_i = \{ 0,1 \}$ and assume the
beneficial variant is present in the population at frequency $p_i$. The
expected change in fitness upon exchange of locus $i$ is therefore
\begin{equation}
\langle \Delta x_i\rangle=\ds \left[ p_i(1-{\bf s}_i)-(1-p_i){\bf
s}_i\right]=\ds (p_i-{\bf s}_i)
\end{equation}
Similarly, the variance of the increment is given by
\begin{equation}
\langle (\Delta x_i-\langle \Delta x_i\rangle)^2\rangle=\ds^2 \left(p_i+{\bf
s}_i -2p_i {\bf s}_i- (p_i-{\bf s}_i)^2\right)=\ds^2 p_i(1-p_i)\, ,
\end{equation}
where we have used ${\bf s}_i={\bf s}_i^2$. Assuming each locus undergoes
exchange with rate $\rec$, the drift and diffusion coefficients of the
fitness $\x$ are given by
\begin{equation}
\langle \Delta \x \rangle = \rec (\X-\bar{X}(t))=r\x \quad \mathrm{and}\quad
\langle (\Delta \x-\langle \Delta \x \rangle)^2 \rangle = \rec\sigma^2
\end{equation}
These diffusion and drift processes are represented by the second and third
terms of   Eq.~(\ref{eq:minimalreckernel}).  The possibility that the novel mutation itself is exchanged into a new genome is described by the first term.

\end{document}